\documentclass[12pt]{article}
\usepackage{amsmath,amsfonts,amssymb}
\newcommand{\be}{\begin{equation}}
\newcommand{\ee}{\end{equation}}
\newcommand{\ba}{\begin{eqnarray}}
\newcommand{\ea}{\end{eqnarray}}

\newcommand{\la}[1]{\label{#1}}

\def\gl#1{(\ref{#1})}

\date{}
\begin{document}
\title{Factorization of non-linear supersymmetry\\ in
one-dimensional Quantum Mechanics. II:\\ proofs of theorems on
reducibility}
\author{
A.V. Sokolov \\
{ \it V.A. Fock Institute of Physics,}\\
{\it Sankt-Petersburg State University}\\ E-mail:
avs{\_}avs@rambler.ru} \maketitle \abstract{In this paper, we
continue to study factorization of supersymmetric (SUSY)
trans\-formations in one-dimensional Quantum Mechanics into chains
of elementary Darboux transformations with nonsingular coefficients.
We define the class of potentials that are invariant under the
Darboux -- Crum transformations and prove a number of lemmas and
theorems substantiating the formulated formerly conjectures on
reducibility of differen\-tial operators for spectral equivalence
transformations. Analysis of the general case is performed with all
the necessary proofs.}

\section{Introduction}

In this work, we present rigorous analysis of factorization of
non-linear supersymmetric (SUSY) Quantum Mechanics \cite{coop} --
\cite{bagsam1} into really irreducible SUSY algebra elements, which
can be used for construction of any polynomial SUSY algebra
\cite{ais} with the help of the chain (ladder) construction
\cite{schr} -- \cite{suku}. From the viewpoint of the Darboux --
Crum (almost) isospectral transformations \cite{crum} --
\cite{matv}, we consider factorization of an intertwining oper\-ator
into a product of differential operators of first or second order
with nonsingular real coefficients such that all the intermediate
Hamiltonians have nonsingular real poten\-tials. The hypothesis on
the existence of such a factorization has been formulated earlier in
\cite{acdi95,samsonov99}.

In \cite{ancan}, it was conjectured that it is possible to dress
(multiply) an intertwining ope\-ra\-tor by a polynomial of the
Hamiltonian preserving the same pair of (almost) isospectral
Hamiltonians and so that the resultant operator may be factorized in
the ladder way into nonsingular real blocks of first order in
derivatives. In this part of the work, conditions for realization of
such a program are found. This part continues the study started in
\cite{andsok1} and we prove here two assertions formulated in
\cite{andsok1} on reducibility of (almost) isospectral
transformations into a chain sequence of irreducible blocks of first
or second order in derivatives:

\renewcommand{\labelenumi}{\rm{(\theenumi)}}
\begin{enumerate}

\item the assertion on reducibility of a nonminimizable intertwining
operator with real spectrum of the matrix $\bf S$, multiplied by an
appropriate polynomial of the Hamiltonian, into (a product of)
intertwining operators of first order (Theorem 2);

\item the assertion on reducibility of a nonminimizable intertwining
operator, whose ma\-trix $\bf S$ may have not only real but also
complex eigenvalues, into (a product of) intertwining operators of
first order and irreducible second-order intertwining oper\-ators of
the I, II and III type \cite{ancan, andsok1} (Theorem 3).

\end{enumerate}

In what follows, we use the class $K$ of potentials $V(x)$ with the
following properties:

\renewcommand{\labelenumi}{\rm{(\theenumi)}}
\begin{enumerate}

\item $V(x)$ is a real-valued function from the class $C_{\Bbb
R}^\infty$;

\item there exist numbers $R_0>0$ and $\varepsilon>0$ (depending on
$V(x)$) such that the inequality $V(x)\geqslant\varepsilon$ takes
place for any $|x|\geqslant R_0$;

\item the functions \be \bigg(\int\limits_{\pm
R_0}^x\sqrt{|V(x_1)|}\,\,dx_1\bigg)^2
\bigg({{|V'(x)|^2}\over{|V(x)|^3}}+{{|V''(x)|}\over{|V(x)|^2}}\bigg)
\ee are bounded for $x\geqslant R_0$ and $x\leqslant -R_0$,
respectively.

\end{enumerate}

In addition, we discuss normalizability and nonnormalizability of
functions at $+\infty$ and/or at $-\infty$, as well as formal
associated functions, which are defined as follows.

A function $f(x)$ is called {\it normalizable at $+\infty$ $($at
$-\infty)$} if there exists a real number $R_+$ ($R_-$) such that
\be \int\limits_{R_+}^{+\infty}|f(x)|^2\,dx<+\infty\qquad
\bigg(\int\limits_{-\infty}^{R_-}|f(x)|^2\,dx<+\infty\bigg).\ee
Otherwise, $f(x)$ is called  {\it nonnormalizable at $+\infty$ $($at
$-\infty)$.}

A function $\psi_{n,i}(x)$ is called a {\it formal associated
function of $i$\hspace{0.2mm}th order} of the Hamiltonian $h$ for a
spectral value $\lambda_n$ if \be
(h-\lambda_n)^{i+1}\psi_{n,i}\equiv0,\qquad
(h-\lambda_n)^{i}\psi_{n,i}\not\equiv 0. \label{canbas1}\ee The term
``formal'' emphasizes that this function is not necessarily
normalizable (not necessarily belongs to $L_2(\Bbb R)$). In
particular, an associated function $\psi_{n,0}(x)$ of zero order is
a formal eigenfunction of $h$ (not necessarily a normalizable
solution of the homogeneous Schr\"odinger equation).

The paper is organized as follows. At first, we present a number of
assertions which clarify basic properties of Hamiltonians with
potentials from the class $K$. These assertions are devoted to (i)
invariance of the class $K$ under intertwining, (ii) asymptotics of
formal associated functions, (iii) properties of a sequence of
formal associated functions under intertwining, and (iv) spectral
properties of intertwined Hamiltonians. Next we prove auxiliary
lemmas on reducibility of operators that intertwine Hamiltonians
with potentials from the class $K$. At last, the main assertions
(Theorems 2 and 3) on reducibility of above-mentioned operators are
stated.

\section{Basic properties of Hamiltonians with \\ \hspace*{10mm} potentials from the class $K$}

Proofs of all the lemmas presented in this section except Lemma 6
and of Theorem 1 are contained in \cite{andcansok,sok1}.

\subsection*{2.1.\quad Invariance of the potential class $K$ under intertwining}

The invariance of the potential class $K$ under intertwining is a
corollary of the following lemma.\\

\noindent {\bf Lemma 1.} {\it Assume that the following conditions
are satisfied:

\renewcommand{\labelenumi}{\rm{(\theenumi)}}
\begin{enumerate}

\item $h^+=-\partial^2+V_1(x)$, $V_1(x)\in K$;

\item $h^-=-\partial^2+V_2(x)$, where the potential $V_2(x)$ is
real-valued and belongs to $C_{\Bbb R}$;

\item $q_N^-h^+=h^-q_N^-$, where $q_N^-$ is a differential operator
of $N$th order with coefficients belonging to $C_{\Bbb R}^2$;

\item each eigenvalue of the matrix ${\bf S}$ for the operator
$q_N^-$ satisfies one of the following conditions: either
$\lambda\leqslant0$ or ${\rm Im}\,\lambda\ne0$.

\end{enumerate}

\noindent Then:

\renewcommand{\labelenumi}{\rm{(\theenumi)}}
\begin{enumerate}

\item $V_2(x)\in K$;

\item coefficients of $q_N^-$ belong to $C_{\Bbb R}^\infty$;

\item $h^+q_N^+=q_N^+h^-$, where $q_N^+=(q_N^-)^t$, and, moreover,
coefficients of $q_N^+$ belong to $C_{\Bbb R}^\infty$ as well.

\end{enumerate}}

\subsection*{2.2.\quad Asymptotics of formal associated functions}

The asymptotic behavior of formal associated functions of a
Hamiltonian $h$ with a potential from the class $K$ is described by
the following lemma.\\

\noindent {\bf Lemma 2.} {\it Assume that the following conditions
are satisfied:\footnote{In what follows, the index $\uparrow$
($\downarrow$) corresponds to upper (lower) signs in the right-hand
sides.}

\renewcommand{\labelenumi}{\rm{(\theenumi)}}
\begin{enumerate}

\item $h=-\partial^2+V(x)$, $V(x)\in K$;

\item $\lambda\in\Bbb C$ and either $\lambda\leqslant0$ or
${\rm{Im}}\,\lambda\ne0$;

\item the branches of the functions $\sqrt{V(x)-\lambda}$ and
$\root4\of{V(x)-\lambda}$ are uniquely defined for $|x|\geqslant
R_0$ by the condition $|\arg[V(x)-\lambda]|<\pi$;

\item  $\xi_{\uparrow\downarrow}(x)=\pm\int\limits_{\pm
R_0}^x\sqrt{|V(x_1)|}\,dx_1$,
$\xi_{\uparrow\downarrow}(x;\lambda)=\pm\int\limits_{\pm
R_0}^x\sqrt{V(x_1)-\lambda}\,dx_1$, $\eta_{\uparrow\downarrow}(x)=$
\linebreak $\pm\int\limits_{\pm R_0}^xdx_1/\sqrt{|V(x_1)|}$.

\end{enumerate}

\noindent Then there exist denumerable sequences:

\renewcommand{\labelenumi}{}
\begin{enumerate}

\item $\varphi_{n,\uparrow\downarrow}(x)$ of formal associated functions
of $h$ for a spectral value $\lambda$ that are nor\-ma\-li\-za\-ble
at $\pm\infty$,

\end{enumerate}

\noindent and

\renewcommand{\labelenumi}{}
\begin{enumerate}

\item $\hat\varphi_{n,\uparrow\downarrow}(x)$ of formal associated
functions of $h$ for a spectral value $\lambda$ that are
non\-nor\-ma\-li\-za\-ble at $\pm\infty$,

\end{enumerate}

\noindent such that:

\renewcommand{\labelenumi}{\rm{(\theenumi)}}
\begin{enumerate}

\item \be
h\varphi_{0,\uparrow\downarrow}=\lambda\varphi_{0,\uparrow\downarrow},
\qquad(h-\lambda)
\varphi_{n,\uparrow\downarrow}=\varphi_{n-1,\uparrow\downarrow},\quad
n=1,2,3,\,\ldots,\la{2.1}\ee \be
h\hat\varphi_{0,\uparrow\downarrow}=\lambda\hat\varphi_{0,\uparrow\downarrow},
\qquad(h-\lambda)
\hat\varphi_{n,\uparrow\downarrow}=\hat\varphi_{n-1,\uparrow\downarrow},\quad
n=1,2,3,\,\ldots;\la{2.2} \ee

\item if $\pm\int\limits_{\pm
R_0}^{\pm\infty}dx_1/\sqrt{|V(x_1)|}<+\infty$, then
\be\varphi_{n,\uparrow\downarrow}(x)={1\over{n!\root
{4\,}\of{V(x)-\lambda}}}
\bigg(\pm{1\over2}\int\limits_{\pm\infty}^x{{dx_1}\over\sqrt{V(x_1)-
\lambda}}\bigg)^ne^{-\xi_{\uparrow\downarrow}(x;\lambda)}\bigg[1+O\bigg({1\over{
\xi_{\uparrow\downarrow}(x)}} \bigg)\bigg],\la{2.3}\ee
\be\hat\varphi_{n,\uparrow\downarrow}(x)={1\over{n!\root
{4\,}\of{V(x)-\lambda}}}
\bigg(\mp{1\over2}\int\limits_{\pm\infty}^x{{dx_1}\over\sqrt{V(x_1)-
\lambda}}\bigg)^ne^{\xi_{\uparrow\downarrow}(x;\lambda)}\bigg[1+O\bigg({1\over{
\xi_{\uparrow\downarrow}(x)}}\bigg)\bigg],\la{2.4}\ee
\be\varphi'_{n,\uparrow\downarrow}(x)=\mp{1\over{n!}}\root4\of{V(x)-\lambda}\bigg(\pm
{1\over2}\int\limits_{\pm\infty}^x{{dx_1}\over\sqrt{V(x_1)-\lambda}}\bigg)^n
e^{-\xi_{\uparrow\downarrow}(x;\lambda)}\bigg[1+O\bigg({1\over{
\xi_{\uparrow\downarrow}(x)}}\bigg)\bigg]\la{2.5}\ee as
$x\to\pm\infty,$ $n=0,1,2,\,\ldots;$

\item if $\pm\int\limits_{\pm
R_0}^{\pm\infty}dx_1/\sqrt{|V(x_1)|}=+\infty$, then
\be\varphi_{n,\uparrow\downarrow}(x)={1\over{n!\root
{4\,}\of{V(x)-\lambda}}} \bigg(\pm{1\over2}\int\limits_{\pm
R_0}^x{{dx_1}\over\sqrt{V(x_1)-
\lambda}}\bigg)^ne^{-\xi_{\uparrow\downarrow}(x;\lambda)}\bigg[1+O\bigg(
{{\ln\eta_{\uparrow\downarrow}(x)}\over{\eta_{\uparrow\downarrow}(x)}}
\bigg)\bigg],\la{2.6}\ee
\be\hat\varphi_{n,\uparrow\downarrow}(x)={1\over{n!\root
{4\,}\of{V(x)-\lambda}}} \bigg(\mp{1\over2}\int\limits_{\pm
R_0}^x{{dx_1}\over\sqrt{V(x_1)-
\lambda}}\bigg)^ne^{\xi_{\uparrow\downarrow}(x;\lambda)}\bigg[1+O\bigg({{\ln
\eta_{\uparrow\downarrow}(x)}\over{\eta_{\uparrow\downarrow}(x)}}
\bigg)\bigg],\la{2.7}\ee
\be\varphi'_{n,{\uparrow\downarrow}}(x)\!=\!\mp{1\over{n!}}\root4\of{V(x)\!-\!\lambda}\bigg(\pm
{1\over2}\!\!\!\int\limits_{\pm
R_0}^x\!\!\!{{dx_1}\over\sqrt{V(x_1)-\lambda}}\bigg)^n
e^{-\xi_{\uparrow\downarrow}(x;\lambda)}\bigg[1+O\bigg({{\ln
\eta_{\uparrow\downarrow}(x)}
\over{\eta_{\uparrow\downarrow}(x)}}\bigg)\bigg]\la{2.8}\ee as $
x\to\pm\infty,$ $n=0,1,2,\,\ldots\,.$

\end{enumerate}}

\vskip0.5pc

\noindent {\bf Corollary 1.} If $V(x)\in K$ and
${\rm{Im}}\,\lambda\ne0$, then, in view of \gl{2.3} and \gl{2.5}
(\gl{2.6} and \gl{2.8}),\footnote{To prove rigorously that the limit
in \gl{W0} is equal to zero, it is sufficient to use, in addition
to~\gl{W0}, the second point of the definition of $K$ and the
estimate ${\rm{Re}}\sqrt{V(x)-\lambda}\geqslant
C_\lambda\sqrt{|V(x)|}$ derived in \cite{sok1}. This estimate is
correct under fixed $V(x)\in K$ and $\lambda$ (with
$\lambda\geqslant0$ or ${\rm{Im}}\,\lambda\ne0$) for some
$C_\lambda>0$ and any $|x|\geqslant R_0$.}
$$\lim\limits_{x\to\pm\infty}[\varphi'_{0,{\uparrow\downarrow}}(x)
\varphi^*_{0,{\uparrow\downarrow}}(x)-
\varphi_{0,{\uparrow\downarrow}}(x)
\varphi'^*_{0,{\uparrow\downarrow}}(x)]=$$
\be\lim\limits_{x\to\pm\infty}
\bigg\{e^{-2{\rm{Re}}\,\xi_{\uparrow\downarrow}(x;\lambda)}
\bigg[\mp{{\root4\of{V(x)-\lambda}}\over{\root4\of{V(x)-\lambda^*}}}(1+o(1))
\pm{{\root4\of{V(x)-\lambda^*}}\over{\root4\of{V(x)-\lambda}}}
(1+o(1))\bigg]\bigg\}=0.\la{W0}\ee Thus, if $V(x)\in K$ and
${\rm{Im}}\,\lambda\ne0$, then the Wronskian $W$ of a function
(denoted below $\varphi$) from ${\rm{ker}}\,(h-\lambda)$ that is
normalizable at one of infinities and of the complex conjugate
function tends to zero at the same infinity; in addition, due to the
monotonicity of $iW$
{$($}$iW'=i(\varphi''\varphi^*-\varphi\varphi''^*)=2{\rm{Im}}\,\lambda
|\varphi|^2${$)$}, the Wronskiasn  does not have zeroes. Let us note
that, in the general case $($contrary to \cite{fern'}$)$, this
statement is not always valid. For example, the Hamiltonian
$$h=-\partial^2-\alpha^2e^{2\beta x}+2\alpha\delta e^{\beta x},\qquad
\alpha\in\Bbb R,\quad \beta>0,\quad\alpha\delta>0,$$ has a formal
eigenfunction for the spectral value
$\lambda=\delta^2-{\beta^2\over4}-i\beta\delta$ of the form
$\varphi(x)=\exp[i{\alpha\over \beta}e^{\beta
x}-(i\delta+{\beta\over 2})x]$. This function tends exponentially to
zero as $x\to+\infty$, but at the same time, the Wronskian
$$W(x)=\varphi'(x)\varphi^*(x)-\varphi(x)\varphi'^*(x)=2i(\alpha-\delta
e^{-\beta x})$$ does not tend to zero as $x\to+\infty$ and has a
real root.

\noindent {\bf Corollary 2.} Under the conditions of Lemma 2, any
formal associated function of $h$ of $n$th order normalizable at
$\pm\infty$, for a spectral value $\lambda$ such that either
$\lambda\leqslant0$ or ${\rm{Im}}\,\lambda\ne0$, can be written in
the form
\be\sum\limits_{j=0}^na_{j,{\uparrow\downarrow}}\varphi_{j,{\uparrow\downarrow}}(x),\qquad
a_{j,{\uparrow\downarrow}}={\rm{Const}},\quad
a_{n,{\uparrow\downarrow}}\ne0,\la{2.9} \ee and any associated
function of $h$ of $n$th order, nonnormalizable at $\pm\infty$, for
the same spectral value $\lambda$ can be presented as follows:
\be\sum\limits_{j=0}^n\big(b_{j,{\uparrow\downarrow}}\varphi_{j,{\uparrow\downarrow}}
(x)+c_{j,{\uparrow\downarrow}}\hat\varphi_{j,{\uparrow\downarrow}}(x)\big),\ee
where
$b_{j,{\uparrow\downarrow}},c_{j,{\uparrow\downarrow}}={\rm{Const}}$
and either $b_{n,{\uparrow\downarrow}}\ne0$ or
$c_{n,{\uparrow\downarrow}}\ne0.$

\subsection*{2.3.\quad Action of an intertwining operator on\\ \hspace*{12.7mm} a sequence
of formal associated functions}

Properties of a sequence of formal associated functions under
intertwining are described by the following lemma.\\

\noindent {\bf Lemma 3.} {\it Assume that:

\renewcommand{\labelenumi}{\rm{(\theenumi)}}
\begin{enumerate}

\item the conditions of Lemma 1 are satisfied;

\item $\varphi_n(x)$, $n=0$, \dots, $M$, is a sequence of formal
associated functions of $h^+$ for a spectral value $\lambda$:
$$h^+\varphi_0=\lambda\varphi_0,\qquad (h^+-\lambda)\varphi_n=
\varphi_{n-1},\quad n=1,\,\ldots,M,$$ where either
$\lambda\leqslant0$ or ${\rm{Im}}\,\lambda\ne0$.

\end{enumerate}

\noindent Then:

\renewcommand{\labelenumi}{\rm{(\theenumi)}}
\begin{enumerate}

\item there is a number $m$, $0\leqslant m\leqslant\min\{M+1,N\}$,
such that
$$ q_N^-\varphi_n\equiv0, \qquad n=0,\,\ldots,\,m-1,$$
and
$$\psi_l=q_N^-\varphi_{m+l},\qquad l=0,\,\ldots,\, M-m,$$ is a sequence of formal
associated functions of $h^-$ for the spectral value $\lambda$:
$$h^-\psi_0=\lambda\psi_0,\qquad (h^--\lambda)\psi_l= \psi_{l-1},\quad
l=1,\,\ldots,\,M-m;$$

\item if a function $\varphi_n(x)$, for a given $0\leqslant
n\leqslant M$, is normalizable at $+\infty$ $($at $-\infty)$, then
$q_N^-\varphi_n$ is normalizable at $+\infty$ $($at $-\infty)$ as
well.

\end{enumerate}}

\vskip0.5pc

\noindent {\bf Corollary 3.} The Hamiltonian $h^+$ is an
intertwining operator for itself, and both eigenvalues of its matrix
${\bf S}$ are zero. Hence, if $\varphi_n(x)$ is normalizable at
$+\infty$ (at $-\infty$), then the functions $\varphi_j(x)$, $j=0$,
\dots $n-1$, are normalizable at $+\infty$ (at $-\infty$) as well.

\noindent {\bf Corollary 4.} Assume that $\varphi^-_{i,j}(x)$ is a
canonical basis in ${\rm{ker}}\,q_N^-$, {\it i.e.}, the matrix ${\bf
S}$ of the operator $q_N^-$ has in this basis the canonical (Jordan)
form:
$$h^+\varphi^-_{i,0}=\lambda_i\varphi^-_{i,0},\,\,\,(h^+-\lambda_i)
\varphi^-_{i,j}=\varphi^-_{i,j-1},\,\,\, i=1,\,\ldots,\, n,\quad
j=1,\,\ldots,\, k_i-1,\,\,\,\sum\limits_{i=1}^n k_i=N.$$ Then there
are numbers $k^+_{i\uparrow}$ and $k^+_{i\downarrow}$ such that
$0\leqslant k^+_{i\uparrow,\downarrow}\leqslant k_i$ and for any
$i$, the functions
$$\varphi^-_{i,j}(x),\qquad j=0,\ldots, k^+_{i{\uparrow,\downarrow}}-1,$$
are normalizable at $\pm\infty$, and the functions
$$\varphi^-_{i,j}(x),\qquad j=k^+_{i{\uparrow,\downarrow}},\ldots,k_i-1,$$
are nonnormalizable at the same $\pm\infty$. Independence of these
numbers $k^+_{i{\uparrow,\downarrow}}$ on a choice of the canonical
basis in the case, where the intertwining operator $q_N^-$ is
nonminimizable, is a corollary of the following lemma.\\

\noindent {\bf Lemma 4.} {\it Assume that:

\renewcommand{\labelenumi}{\rm{(\theenumi)}}
\begin{enumerate}

\item the conditions of Lemma 1 are satisfied;

\item $q_N^-$ is non\-mi\-ni\-mi\-za\-ble.

\end{enumerate}

\noindent Then any two formal associated functions of $h^+$ of the
same order for the same spectral value $\lambda$ when being elements
of ${\rm{ker}}\,q_N^-$ are either simultaneously normalizable at
$+\infty$ or simultaneously nonnormalizable at $+\infty$. The same
fact takes place at $-\infty$.}

\subsection*{2.4.\quad Statements on spectra of intertwined Hamiltonians}

The following Lemma 5 clarifies an interrelation between the
behavior at $\pm\infty$ of elements of canonical bases for mutually
transposed
intertwining operators.\\

\noindent {\bf Lemma 5.} {\it Assume that:

\renewcommand{\labelenumi}{\rm{(\theenumi)}}
\begin{enumerate}

\item the conditions of Lemma 1 are satisfied;

\item $q_N^-$ is nonminimizable;

\item $k_i$ is algebraic multiplicity of an eigenvalue $\lambda_i$ of
the matrix $\bf S$ of the operator $q_N^-$;

\item $\{\varphi^-_{i,j}\}$ and
$\{\varphi^+_{i,j}\}$ are canonical bases of ${\rm{ker}}\,q_N^-$ and
${\rm{ker}}\,q_N^+$, respectively:
$$h^\pm\varphi^\mp_{i,0}\!=\!\lambda_i\varphi^\mp_{i,0},\quad(h^\pm\!-\!\lambda_i)
\varphi^\mp_{i,j}\!=\!\varphi^\mp_{i,j-1},\quad i\!=\!1,\,\ldots,\,
n,\,\,\, j\!=\!1,\,\ldots,\, k_i\!-\!1,\quad\sum\limits_{i=1}^n
k_i\!=\!N.$$

\end{enumerate}

\noindent Then for any $i$ and $j$, the function
$\varphi^-_{i,j}(x)$ is normalizable (nonnormalizable) at $+\infty$
if and only if $\varphi^+_{i, k_i-j-1}(x)$ is nonnormalizable
(normalizable) at $+\infty$. The same fact takes place at
$-\infty$.}

\newpage

\noindent {\bf Corollary 5.} In the family $\varphi^-_{i,j}(x)$
($\varphi^+_{i,j}(x)$) (with a fixed $i$) only the function
$\varphi^-_{i,{k_i-1}}(x)$ ($\varphi^+_{i,{k_i-1}}(x)$) may be
nonnormalizable at both infinities. Thus, in view of Corollary 3,
one of the numbers $k^+_{i\uparrow,\downarrow}$ (with a fixed $i$)
is not less than $k_i-1$ and the other one is not greater than~1.
Accordingly, the functions $\varphi^-_{i,j}(x)$
($\varphi^+_{i,j}(x)$), $j=1$, \dots, $k_i-2$, are normalizable at
one of infinities (the same for any $j$) and are nonnormalizable at
the other infinity. Moreover, if the functions $\varphi^-_{i,j}(x)$
are normalizable at $+\infty$ (at $-\infty$), then the functions
$\varphi^+_{i,j}(x)$ are normalizable at $-\infty$ (at $+\infty$).

\noindent {\bf Corollary 6.} If $\varphi^-_{i,0}(x)$
($\varphi^+_{i,0}(x)$) is nonnormalizable at both infinities, then
$k_i=1$.

\noindent {\bf Corollary 7.} If both functions $\varphi^-_{i,0}(x)$
and $\varphi^+_{i,0}(x)$ are normalizable at both infinities, then
$k_i\geqslant2$.

\noindent {\bf Corollary 8.} If ${\rm{Im}}\,\lambda_i\ne0$, then, in
view of Corollary 3, the functions $\varphi^-_{i,j}(x)$
($\varphi^+_{i,j}(x)$), $j=0$,~\dots, $k_i-1$, are normalizable at
one of infinities (the same for any $j$) and nonnormalizable at the
other infinity. Moreover, if the functions $\varphi^-_{i,j}(x)$ are
normalizable at $+\infty$ (at $-\infty$), then functions
$\varphi^+_{i,j}(x)$ are
normalizable at $-\infty$ (at $+\infty$).\\

The following theorem indicates an interrelation between spectra of
intertwined Hamiltonians and the behavior at $\pm\infty$ of elements
of a canonical basis in the intertwining ope\-ra\-tor kernel.\\

\noindent {\bf Theorem 1} (Index Theorem). {\it Assume that the
conditions of Lemma 5 are satisfied. Set $\nu_\pm(\lambda)=1$ if
$\lambda$ is an eigenvalue of $h^\pm$ and $\nu_\pm(\lambda)=0$
otherwise. As well set $n_\pm(\lambda)= n_0(\lambda)=0$ if $\lambda$
is not an eigenvalue of the matrix $\bf S$ of the operator $q_N^-$.
If $\lambda=\lambda_i$ (where $\lambda_i$ is an eigenvalue of the
matrix $\bf S$ of the operator $q_N^-$), let $n_\pm(\lambda_i)$ be
the number of functions from the family $\varphi^\mp_{i,j}(x)$,
$j=0$, \dots, $k_i-1$, that are normalizable at both infinities and
let $n_0(\lambda_i)$ be the number of functions from the family
$\varphi^\mp_{i,j}(x)$, $j=0$, \dots, $k_i-1$, that are normalizable
only at one of infinities. Then for any $\lambda$ such that either
$\lambda\leqslant0$ or ${\rm {Im}}\,\lambda\ne0$ the equality
\be\nu_+(\lambda)-n_+(\lambda)=\nu_-(\lambda)-n_-(\lambda)\la{nnnn}\ee
takes place. Moreover, if $n_0(\lambda)>0$ for some $\lambda$, then
$$\nu_+(\lambda)-n_+(\lambda)=\nu_-(\lambda)-n_-(\lambda)=0$$
for this $\lambda$.}\\

The next lemma indicates an interrelation between the inclusion of a
bound state wave function of a Hamiltonian to the kernel of an
intertwining operator and the inclusion of the energy of the same
state to the spectrum of the matrix $\bf S$ of the
considered in\-ter\-twining operator.\\

\noindent {\bf Lemma 6.} {\it If the conditions of Lemma 1 are
satisfied, then a wave function of a bound state of $h^\pm$ belongs
to ${\rm{ker}}\,q_N^\mp$ if and only if the energy of this bound
state is contained in the spectrum of the matrix $\bf S$
of the operator $q_N^\mp$.}\\

\noindent{\bf Proof.} We only consider the case of $h^+$ and $q_N^-$
since the case of $h^-$ and $q_N^+$ is treated similarly.

NECESSITY. Assume that $h^+$ has a bound state with energy $E$ which
is described by a wave function $\varphi(x)$ and that, in addition,
$q_N^-\varphi=0$. We claim that $E$ belongs to the spectrum of the
matrix $\bf S$ of the operator $q_N^-$. Let $\lambda_i$ be an
eigenvalue of the matrix $\bf S$ of the operator $q_N^-$ of
algebraic multiplicity $k_i$, $i=1$, \dots, $n$, so that
$k_1+\ldots+k_n=N$. By Theorem 1 of \cite{andsok1},
\be0=q_N^+q_N^-\varphi=\prod\limits_{i=1}^n(h^+-\lambda_i)^{k_i}\varphi=
\prod\limits_{i=1}^n(E-\lambda_i)^{k_i}\varphi,\ee from which it
follows that $E$ belongs to the spectrum of the matrix $\bf S$ of
the operator $q_N^-$.

SUFFICIENCY. We assume now that $E$ belongs to the spectrum of the
matrix $\bf S$ of the operator $q_N^-$. Let us show that
$q_N^-\varphi=0$. Let \be q_N^-=p_M^-P(h^+),\ee where $P(h^+)$ is a
polynomial and $p_M^-$ is a nonminimizable operator which
intertwines $h^+$ and $h^-$ ($p_M^-h^+=h^-p_M^-$). If $E$ is a zero
of $P$, then the statement is proved. Let us proceed to to the case
$P(E)\ne0$. In this case, $E$ belongs to the spectrum of the matrix
$\bf S$ of the operator $p_M^-$, because by Theorem 1 of
\cite{andsok1}, the spectrum of the matrix $\bf S$ of the operator
$q_{N}^-$ coincides with the set of zeroes of the polynomial
$P_N(h^+)=q_N^+q_N^-$, the spectrum of the matrix $\bf S$ of the
operator $p_M^-$ coincides with the set of zeroes of the polynomial
$P_M(h^+) =p_M^+p_M^-$, $p_M^+=(p_M^-)^t$ and \be
P_N(h^+)=q_N^+q_N^-=P(h^+)p_M^+p_M^-P(h^+)= P^2(h^+)P_M(h^+).\ee Let
$q_N^-\varphi\ne0$. Then $p_M^-\varphi\ne0$ as well, since otherwise
$q_{N}^-\varphi=P(h^-)p_M^-\varphi=0$. By Lem\-ma~3, $p_M^-\varphi$
is an eigenfunction of $h^-$ that belongs to ${\rm{ker}}\,p_M^+$
since $p_M^+p_M^-\varphi= P_M(E)\varphi=0$. The latter fact
contradicts statement \gl{nnnn} of Theorem 1, because in the case
under consideration, $\nu_+(E)=\nu_-(E)=1$, and by Lemmas 4 and 5,
$n_+(E)=0$ and $n_-(E)=1$ (in the considered case, $n_\pm(E)$
corresponds to $p_M^-$ and not to
$q_N^-$). Lemma 6 is proved.\\

\noindent {\bf Corollary 9.} By Lemmas 5 and 6, $h^\pm$ has a bound
state at a level $E=\lambda_i$ if and only if the function
$\varphi^+_{i,k_i-1}(x)$ ($\varphi^-_{i,k_i-1}(x)$) is
nonnormalizable at both infinities.

\noindent {\bf Corollary 10.} Assume that at least one of the
coefficients of $q_N^\mp$ has a nontrivial imaginary part and that
$k_N^\mp$ and $p_M^\mp$ are differential operators with real-valued
coefficients such that $q_N^\mp=k_N^\mp+ ip_M^\mp$. Then (see
\cite{ansok}) the operators $k_N^\mp$ and $p_M^\mp$ intertwine the
same Hamiltonians as $q_N^\mp$. Moreover, since a wave function of a
bound state can be chosen real-valued, any wave function of a bound
state that belongs to ${\rm{ker}}\,q_N^\mp$ belongs to
${\rm{ker}}\,k_N^\mp$ and ${\rm{ker}}\,p_M^\mp$ as well. Hence, any
eigenvalue of the matrix $\bf S$ of the operator $q_N^\mp$, which is
the energy of a bound state of $h^\pm$, belongs to the spectra of
the matrices $\bf S$ of the operators $k_N^\mp$ and $p_M^\mp$ as
well.

\section{Lemmas on partial reducibility of\\ \hspace*{10mm} intertwining operators}

\noindent {\bf Lemma 7.} {\it Assume that:

\renewcommand{\labelenumi}{\rm{(\theenumi)}}
\begin{enumerate}

\item the conditions of Lemma 5 are satisfied;

\item all coefficients of $q_N^-$ are real-valued;

\item ${\rm{Im}}\,\lambda_l\ne0$.

\end{enumerate}

\noindent Then $q_N^-$ can be represented as the product of two
intertwining operators $k_{N-2}^-$ and $p_2^-$, so that:

\renewcommand{\labelenumi}{\rm{(\theenumi)}}
\begin{enumerate}

\item \be q_N^-=k_{N-2}^-p_2^-,\qquad h_0p_2^-=p_2^-h^+,\qquad
h^-k_{N-2}^-=k_{N-2}^-h_0,\la{fakt1}\ee where $h_0$ is the
Hamiltonian with the potential from $K$;

\item $p_2^-$ is the really irreducible intertwining operator of
second order of the I type with real-valued coefficients from
$C_{\Bbb R}^\infty$, and the spectrum of the matrix $\bf S$ of the
operator $p_2^-$ consists of $\lambda_l$ and~$\lambda_l^*$;

\item $k_{N-2}^-$ is the intertwining operator of $(N-2)$th order
with real-valued coefficients from~$C_{\Bbb R}^\infty$.

\end{enumerate}}

\vskip0.5pc

\noindent{\bf Proof.} Taking into account reality of coefficients of
$q_N^-$, we assume, without loss of generality, that a basis
$\{\varphi^-_{i,j}\}$ in the kernel of $q_N^-$ is chosen so that the
functions $\varphi^-_{i,j}$, corresponding to real $\lambda_i$, are
real-valued and the functions $\varphi^-_{i,j}$ and
$\varphi^-_{k,j}$, corresponding to complex conjugated numbers
$\lambda_i$ and $\lambda_k=\lambda_i^*$, are related by
$\varphi^-_{k,j}=\varphi^{-*}_{i,j}$.

Using the procedure described in Lemma 1 of \cite{ansok}, one can
represent $q_N^-$ in the form \be q_N^-=k_{N-2}^-p_2^-,\la{fakt5}\ee
where $p_2^-$ is the differential operator of second order whose
kernel basis consists of $\varphi^-_{l,0}(x)$ and
$\varphi^{-*}_{l,0}(x)$, and $k_{N-2}^-$ is the differential
operator of $(N-2)$th order whose kernel basis consists of
$p_2^-\varphi^-_{i,j}$, with the exception of $p_2^-\varphi^-_{l,0}$
and $p_2^-\varphi^{-*}_{l,0}$. Moreover, by the above-mentioned
lemma, $p_2^-$ and $k_{N-2}^-$ intertwine $h^+$ and $h^-$,
respectively, with certain Hamiltonian $h_0$, so that equalities
\gl{fakt1} hold.

Let us denote the Wronskians of elements of the above-mentioned
bases in ${\rm{ker}}\,k_{N-2}^-$ and ${\rm{ker}}\,p_2^-$ by $W_k(x)$
and $W_p(x)$, respectively. Then, by formula (11) of \cite{andsok1},
the potential $V_0(x)$ of the Hamiltonian $h_0$ is related to
$V_{1,2}(x)$ by the following equalities: \be V_0(x)=V_1(x)-2[\ln
W_p(x)]'', \qquad V_2(x)=V_0(x)-2[\ln W_k(x)]''.\la{vvvv}\ee In view
of Corollary 8 the function $\varphi^-_{l,0}(x)$ is normalizable at
one of infinities. Thus, by Corollary 1, the Wronskian $W_p(x)$ does
not have zeroes.  We derive from this fact and from the inclusion
$\varphi^-_{l,0}(x)\in C^\infty_\Bbb R$ that $V_0(x)$ (see
\gl{vvvv}) and coefficients of \be p_2^-={1\over{
W_p(x)}}\begin{vmatrix}\varphi^{-*}_{l,0}(x)&\varphi^{-\prime*}_{l,0}(x)&
\varphi^{-\prime\prime*}_{l,0}(x)\\ \varphi^-_{l,0}(x)&
\varphi^{-\prime}_{l,0}(x)&\varphi^{-\prime\prime}_{l,0}(x)\\
1&\partial&\partial^2\end{vmatrix}\la{q2}\ee belong to
$C^\infty_\Bbb R$. Moreover, coefficients of $p_2^-$ are real since
complex conjugation of these coefficients is equivalent to
permutations of two lines in both determinants in \gl{q2}. The fact
that $V_0(x)$ is real-valued follows from \gl{vvvv} and from the
fact that $W_p(x)$ is evidently purely imaginary. The inclusion of
$V_0(x)$ into $K$ follows from the statements proved above and from
Lemma 1.

The absence of zeroes for $W_k(x)$ follows from the infinite
smoothness of $V_0(x)$ and $V_2(x)$ and from the fact that the
general solution of \gl{vvvv} has the form
$$W_k(x)=C_1\exp\Big\{C_2x+{1\over2}\int\limits_0^xdx_1\int\limits_0^{x_1}
dx_2\,[V_0(x_2)-V_2(x_2)]\Big\},$$ where $C_1\ne0$ and $C_2$ are
constants. The infinite smoothness of coefficients of $k_{N-2}^-$
follows from the absence of zeroes for $W_k(x)$, from the infinite
smoothness of $W_k(x)$ (the Wronskian of functions from
$C^\infty_\Bbb R$), and from the formula for $k_{N-2}^-$ similar to
\gl{q2}. The fact that coefficients of $k_{N-2}^-$ are real-valued
is an obvious corollary of the fact that coefficients of $q_N^-$ and
$p_2^-$ are real-valued. Lemma 7 is proved.\\

\noindent {\bf Lemma 8.} {\it Assume that:

\renewcommand{\labelenumi}{\rm{(\theenumi)}}
\begin{enumerate}

\item the conditions of Lemma 5 are satisfied;

\item $\lambda_M$ is the least real eigenvalue of the matrix $\bf S$
of the operator $q_N^-$;

\item $\lambda_M$ is situated below the energy of the ground state of
$h^-$.

\end{enumerate}

\noindent Then $q_N^-$ can be factorized into the product of two
intertwining operators $k_{N-1}^-$ and $p_1^-$, so that:

\renewcommand{\labelenumi}{\rm{(\theenumi)}}
\begin{enumerate}

\item \be q_N^-=k_{N-1}^-p_1^-,\qquad h_0p_1^-=p_1^-h^+,\qquad
h^-k_{N-1}^-=k_{N-1}^-h_0,\la{fakt1'}\ee where $h_0$ is the
Hamiltonian with the potential from $K$;

\item $p_1^-$ is the intertwining operator of first order with
real-valued coefficients from $C_{\Bbb R}^\infty$, and its matrix
$\bf S$ consists of $\lambda_M$;

\item $k_{N-1}^-$ is the intertwining operator of $(N-1)$th order with
coefficients from $C_{\Bbb R}^\infty$;

\item if $\lambda_M$ is (is not) the energy of a bound state of $h^+$,
then an element of a basis in ${\rm{ker}}\,p_1^-$ is normalizable at
both infinities (at one of infinities only).

\end{enumerate}

\noindent If coefficients of $q_N^-$ are real-valued, then
coefficients of
$k_{N-1}^-$ are real-valued as well.}\\

\noindent{\bf Proof.} By the conditions of our lemma and by Lemmas 3
and 6, the number $\lambda_M$ is not situated above the energy of
the ground state of $h^+$. By our conditions and by Lem\-ma~5, the
function $\varphi^-_{M,0}(x)$ cannot be nonnormalizable at both
infinities. At the same time, consider a formal eigenfunction of
$h^+$ that is normalizable at least at one of infinities and
corresponds to a spectral value that is not situated above the
energy of the ground state. Such a function has no zeroes and may
differ from a real-valued function (if such a difference exists) by
a constant factor only. Hence, the operator \be
p_1^-=\partial-{{\varphi^{-\prime}_{M,0}}\over{\varphi^-_{M,0}}}\ee
has real-valued coefficients from $C_{\Bbb R}^\infty$. In accordance
with the procedure described in Lemma 1 of \cite{ansok}, this
operator can be separated from $q_N^-$, so that the equalities
\gl{fakt1'} are valid, where $k_{N-1}^-$ is the intertwining
operator of $(N-1)$th order with coefficients from~$C_{\Bbb
R}^\infty$, and $h_0$ is the Hamiltonian whose potential $V_0(x)$,
by the relation (11) of \cite{andsok1}, is equal to \be
V_0(x)=V_1(x)-2[\ln \varphi^-_{M,0}(x)]''.\la{vm}\ee We deduce that
$V_0(x)$ is real-valued and infinitely smooth from relation \gl{vm},
from the absence of zeroes for $\varphi^-_{M,0}(x)$, from the
inclusions $V_1(x)\in K$ and $\varphi^-_{M0}(x)\in C_{\Bbb
R}^\infty$, and from the proportionality of $\varphi^-_{M,0}(x)$ to
a real-valued function. The inclusion of $V_0(x)$ into the class $K$
follows from the statements proven above and from Lemma 1. The
fourth statement of the lemma follows from the normalizability of
$\varphi^-_{M,0}(x)$ at least at one of infinities and from Lemma 6.
If coefficients of $q_N^-$ are real-valued, then coefficients of
$k_{N-1}^-$ are obviously real-valued as well. Lemma 8 is proved.\\

\noindent {\bf Lemma 9.} {\it Assume that:

\renewcommand{\labelenumi}{\rm{(\theenumi)}}
\begin{enumerate}

\item the potential of the Hamiltonian $h^+$ belongs to $K$; the
potential of the Hamiltonian $h_1$ is real-valued and belongs to
$C^1_{\Bbb R}$; the potential of the Hamiltonian $h^-$ is
real-valued and belongs to $C_{\Bbb R}$;

\item $\varphi_0(x)$ is a wave function of the ground state of $h^+$,
so that \be h^+\varphi_0=E_{0+}\varphi_0,\qquad E_{0+}\leqslant0;\ee
the Hamiltonians $h^+$ and $h_1$ are intertwined by the operator
$p_{11}^-=\partial-\varphi'_0/ \varphi_0$, so that \be
p_{11}^-h^+=h_1p_{11}^-;\ee

\item $\psi(x)$ is a function that is normalizable at one of infinities only
and belongs to ${\rm{ker}}\,(h_1-\lambda)$, $\lambda<E_{0+}$; the
Hamiltonians $h_1$ and $h^-$ are intertwined by the ope\-ra\-tor
$k_{11}^-=\partial-\psi'/\psi$, so that \be
k_{11}^-h_1=h^-k_{11}^-.\ee

\end{enumerate}

\noindent Then:

\renewcommand{\labelenumi}{\rm{(\theenumi)}}
\begin{enumerate}

\item the potentials of $h^-$ and $h_1$ belong to $K$; coefficients
of $p_{11}^-$ and $k_{11}^-$ are real-valued and belong to $C_{\Bbb
R}^\infty$;

\item the function $(p_{11}^-)^t\psi$ does not have zeroes, belongs to
${\rm{ker}}\, (h^+-\lambda)$, and is normalizable at one of
infinities only (the same as $\psi$);

\item the operator \be
p_{12}^-=\partial-{{(p_{11}^+\psi)'}\over{p_{11}^+\psi}},\qquad
p_{11}^+=(p_{11}^-)^t\ee has real-valued cofficients from $C_{\Bbb
R}^\infty$ and intertwines $h^+$ with the Hamiltonian
$h_2=\lambda+p_{12}^-(q_{12}^-)^t$, so that \be
p_{12}^-h^+=h_2p_{12}^-;\ee the potential of $h_2$ belongs to $K$;
the matrix $\bf S$ of the operator $p_{12}^-$ consists of $\lambda$;

\item $p_{12}^-\varphi_0$ is a wave function of the ground state of
$h_2$ with the energy $E_{0+}$;

\item the operator \be
k_{12}^-=\partial-{{(p_{12}^-\varphi_0)'}\over{p_{12}^-\varphi_0}}\ee
has real-valued coefficients from $C_{\Bbb R}^\infty$ and
intertwines $h_2$ with $h^-$, so that \be
k_{12}^-h_2=h^-k_{12}^-;\ee the matrix $\bf S$ of the operator
$k_{12}^-$ consists of $E_{0+}$;

\item the equality \be k_{11}^-p_{11}^-=k_{12}^-p_{12}^-\ee holds.

\end{enumerate}}

\vskip0.5pc

Lemma 9 follows trivially from Lemmas 1 and 3, Theorem 1 of
\cite{andsok1}, standard construction which describes intertwining
of Hamiltonians by operators of first order \cite{coop} --
\cite{bagsam1}, \cite{schr} -- \cite{suku}, \cite{ancan}, and
elementary information on zeroes of formal
eigenfunctions of a Hamiltonian \cite{shubin}.\\

\noindent {\bf Lemma 10.} {\it Assume that:

\renewcommand{\labelenumi}{\rm{(\theenumi)}}
\begin{enumerate}

\item the conditions of Lemma 1 are satisfied with $N=3$;

\item $q_3^-$ is nonminimizable, and its coefficients are real-valued;

\item $\lambda$ is the least real eigenvalue of the matrix $\bf S$ of
the operator $q_3^-$.

\end{enumerate}

\noindent Then there exist intertwining operators $p_1^\pm$ and
$k_{1}^\pm$ of first orders and $p_{2}^\pm$ and $k_{2}^\pm$ of
second orders such that:

\renewcommand{\labelenumi}{\rm{(\theenumi)}}
\begin{enumerate}

\item $p_{1}^\pm$, $k_{1}^\pm$, $p_{2}^\pm$ and $k_{2}^\pm$ have
real-valued coefficients from $C^\infty_{\Bbb R}$;

\item \be p_1^+=(p_1^-)^t,\qquad k_1^+=(k_1^-)^t,\qquad
p_2^+=(p_2^-)^t,\qquad k_2^+=(k_2^-)^t;\ee

\item the matrices $\bf S$ of the operators $p_{1}^\pm$ and
$k_{1}^\pm$ consist of $\lambda$;

\item \be q_3^-=k_{2}^-p_{1}^-=k_{1}^-p_{2}^-,\qquad
q_3^+=p_{1}^+k_{2}^+=p_{2}^+k_{1}^+,\ee and the potentials of
intermediate Hamiltonians that correspond to these factorizations
belong to $K$;

\item if the algebraic multiplicity of $\lambda$ in the spectrum of the
matrix $\bf S$ of the operator $q_3^-$ is equal to one, then an
element of a basis in ${\rm{ker}}\,p_{1}^-$ is normalizable
(nonnormalizable) at $+\infty$ if and only if an element of a basis
in ${\rm{ker}}\,k_{1}^-$ is normalizable (non\-nor\-ma\-li\-za\-ble)
at $+\infty$; the same fact is true at $-\infty$; the same facts
take place for $p_{1}^+$ and~$k_{1}^+$.

\end{enumerate}}

\vskip0.5pc

\noindent{\bf Proof.} The first four statements of the lemma follow
from Theorem 3 of \cite{andsok1} and from Lemma 1. In the proof of
the fifth statement of the lemma, we consider the case of $p_{1}^-$,
$k_{1}^-$ and $x\to+\infty$ only, since the remaining cases can be
examined analogously. Let $\varphi(x)$ be an element of a basis in
${\rm{ker}}\,p_{1}^-$. As the matrix $\bf S$ of the operator
$p_{1}^-$ consists of $\lambda$, so $h^+\varphi=\lambda\varphi$. Let
also $\lambda_1$ and $\lambda_2$ be the remaining two eigenvalues of
the matrix $\bf S$ of the operator $q_3^-$ as well as $\varphi_1(x)$
and $\varphi_2(x)$ be the remaining two elements of a canonical
basis in ${\rm{ker}}\,q_3^-$, where, by condition,
$\lambda\ne\lambda_{1,2}$. As $q_3^-=k_{1}^-p_{2}^-$,
$q_3^-\varphi=0$, and the basis in ${\rm{ker}}\,p_{2}^-$ consists of
$\varphi_1(x)$ and $\varphi_2(x)$, so \be
\psi=p_{2}^-\varphi\not\equiv0\la{zv1}\ee is the only element of the
basis in $\ker k_{1}^-$. On the other hand, by Theorem 1 of
\cite{andsok1}, \be
p_{2}^+\psi=p_{2}^+p_{2}^-\varphi=(h^+-\lambda_1)(h^+-\lambda_2)\varphi=
(\lambda-\lambda_1)(\lambda-\lambda_2)\varphi\not\equiv0.\la{zv2}\ee
It follows from equalities \gl{zv1} and \gl{zv2} and from Lemma 3
that the normalizability of $\varphi(x)$ at $+\infty$ is equivalent
to the normalizability of $\psi(x)$ at $+\infty$. Lemma 10 is
proved.

\section{Theorems on complete reducibility of\\ \hspace*{10mm} intertwining operators}

\noindent {\bf Theorem 2} (on reducibility of ``dressed''
nonminimizable intertwining operators).

\noindent{\it Assume that the following conditions are satisfied:

\renewcommand{\labelenumi}{\rm{(\theenumi)}}
\begin{enumerate}

\item $h^+=-\partial^2+V_1(x)$, $V_1(x)\in K$;

\item $h^-=-\partial^2+V_2(x)$, where the potential $V_2(x)$ is real-valued
and belongs to $C_{\Bbb R}$;

\item $h^+$ and $h^-$ are intertwined by a nonminimizable
differential operator $q_N^-$ of $N$th order with coefficients from
$C^2_{\Bbb R}$, so that \be q_N^-h^+=h^-q_N^-;\ee

\item the algebraic multiplicity of $\lambda_i$, the $i$\hspace{0.3mm}th
eigenvalue of the matrix $\bf S$ for the operator $q_N^-$, is equal
to $k_i$, $i=1$, \dots, $n$, so that $k_1+\dots+k_n =N$; all of the
numbers $\lambda_i$ are real and satisfy the inequalities \be
0\geqslant\lambda_1>\lambda_2>\ldots>\lambda_n;\ee

\item $\Lambda$ is the spectrum of the matrix $\bf S$ of the
operator $q_N^-$;

\item $E_{i\pm}$, $i=0$, $1$, $2$, \dots is the energy of the $i$\hspace{0.3mm}th
(from below) bound state of $h^\pm$, $N^\pm$ is the number of bound
states of $h^\pm$ whose energies are included into $\Lambda$, and
$N_\pm$ is the number of bound states of $h^\pm$ with energies not
exceeding $\lambda_1$;

\item \be
P_\pm(E)=\prod\limits_{E_{i\pm}<\lambda_1,\,E_{i\pm}\not\in\Lambda}
(E_{i\pm}-E).\ee

\end{enumerate}

\noindent Then:

\renewcommand{\labelenumi}{\rm{(\theenumi)}}
\begin{enumerate}

\item $V_2(x)\in K$; coefficients of $q_N^-$ belong to
$C^\infty_{\Bbb R}$ and are real-valued; $q_N^+=(q_N^-)^t$ has
real-valued coefficients from $C^\infty_{\Bbb R}$ and intertwines
$h^+$ and $h^-$, so that \be h^+q_N^+=q_N^+h^-;\la{splet}\ee

\item $P_+(E)\equiv P_-(E)$; the degree of $P_\pm(E)$ is equal to
$N_+-N^+=N_--N^-$;

\item the operator $q_N^\mp P_\pm(h^\pm)$ intertwines $h^+$ and
$h^-$ and can be represented as the product of $N+N_++N_--N^+-N^-$
intertwining operators of first order with real-valued coefficients
from $C^\infty_{\Bbb R}$, so that:

\renewcommand{\labelenumii}{\rm{(\theenumii)}}
\begin{enumerate}

\item potentials of all the intermediate Hamiltonians belong to
$K$;

\item the eigenvalue of the matrix $\bf S$ of the $l$\hspace{0.3mm}th operator
(from the right) in the fac\-to\-ri\-za\-ti\-on under consideration
is equal to $E_{l-1,\pm}$, $l=1$, \dots, $N_\pm$, and an element of
the kernel of this operator is normalizable at both infinities;

\item the eigenvalue of the matrix $\bf S$ of the $l$\hspace{0.3mm}th operator
(from the left) in the fac\-to\-ri\-za\-ti\-on under consideration
is equal to $E_{l-1,\mp}$, $l=1$, \dots, $N_\mp$, and an element of
the kernel of this operator is nonnormalizable at both infinities;

\item the set of eigenvalues of the matrices $\bf S$ for operators
from the $(N_\pm+1)$th to the $(N_\pm+N-N^+-N^-)$th one (from the
right) in the factorization under consideration coincides
with\footnote{In this formula, one has to take into account
multiplicities of eigenvalues as follows: if $\lambda$ is contained
in $\Lambda$ with algebraic multiplicity $K_1$, in $\{E_{i+}\}$ with
multiplicity $K_2$, and in $\{E_{i-}\}$ with multiplicity $K_3$
(obviously, $K_2$ and $K_3$ can take values 0 and 1 only), then the
value $\lambda$ is contained in
$\Lambda\setminus(\{E_{i+}\}\cup\{E_{i-}\})$ with multiplicity
$K_1-K_2-K_3$ if $K_1>K_2+K_3$ or is not contained if $K_1\leqslant
K_2+K_3$.} $\Lambda\setminus(\{E_{i+}\}\cup\{E_{i-}\})$. In
addition, the eigenvalue of the matrix $\bf S$ for an operator of
this group does not decrease as the number of the operator increases
(from the right to left); a basis element of the kernel of any
operator of this group is normalizable at one of the infinities
only.

\end{enumerate}\end{enumerate}}

\vskip0.5pc

\noindent{\bf Proof.} The first statement of the theorem with the
exception of reality of $q_N^\pm$ coefficients follows from Lemma 1.
The fact, that coefficients of $q_N^\pm$ are real-valued, will be
proved below.

To prove the second statement, it is obviously sufficient to show
that if $E_{i\pm}<\lambda_1$ and $E_{i\pm}\not\in\Lambda$, then
there exist an $E_{j\mp}$ such that $E_{j\mp}=E_{i\pm}$. The latter
fact follows from Lemmas 3 and 6. Thus, the second statement is
proved.

Intertwining of $h^+$ and $h^-$ by the operators $q_N^\mp
P_\pm(h^\pm)$ is evident.

By the definition of $P_\pm$ and Lemma 6, the kernel
${\rm{ker}}\,(q_N^\mp P_\pm(h^\pm))$ contains wave func\-tions of
$N_\pm$ lower bound states of $h^\pm$. Moreover, in view of the
nonminimizability of $q_N^\mp$, the canonical basis in
${\rm{ker}}\,(q_N^\mp P_\pm(h^\pm))$ can be chosen to contain all
these wave functions. Using the standard procedure described in
Lemma 1 of \cite{ansok}, one can separate successively from the
right $q_N^\mp P_\pm(h^\pm)$ intertwining operators of first orders
whose kernels bases consist of ground state wave functions of
$h^\pm$ or of the corresponding intermediate Hamiltonians. In
addition, it is easy to verify that coefficients of separated
intertwining operators are real-valued and infinitely smooth, and
that potentials of intermediate Hamiltonians be\-long to $K$ by
induction with the help of the reasoning from the proof of Lemma~8.
Thus, statement 3(b) is proved, and the ground state of the last
intermediate Hamiltonian $h_0^\pm$ is situated above $\lambda_1$.

Let $k_\mp$ be the remainder of $q_N^\mp P_\pm(h^\pm)$ after
realization of all the above-mentioned se\-pa\-ra\-ti\-ons. This
operator intertwines $h_0^\pm$ and $h^\mp$, so that $k_\mp
h_0^\pm=h^\mp k_\mp$ and $h_0^\pm k_\mp^t=k_\mp^t h^\mp$. Thus, by
Lemma 3 and due to the absence of energy levels of $h_0^\mp$ that
are not situated above $\lambda_1$, wave functions of $N_\mp$ lower
bound states of $h^\mp$ belong to ${\rm{ker}}\,k_\mp^t$. On the
other hand, the nonminimizability of $q_N^\mp$, Theorem 2 of
\cite{andsok1}, and the rule of transformation of the Jordan form of
the matrix $\bf S$ of an intertwining operator under separation from
it an intertwining operator of first order (see Lemma 1 of
\cite{ansok}) imply that the operators $k_\mp$ and thereby $k_\mp^t$
are nonminimizable. Hence, wave functions of $N_\mp$ lower bound
states of $h^\mp$ belong to a canonical basis in
${\rm{ker}}\,k_\mp^t$. Using the same separation procedure as above
and taking into account that a product of elements of bases in
${\rm{ker}}\, (\partial-\chi(x))$ and
${\rm{ker}}\,(\partial-\chi(x))^t$ is constant, we establish
statement 3(c).

Let us denote by $r_\mp$ the remainder of $k_\mp$ after separation
of the operators mentioned in statement 3(c). This operator is
nonminimizable again (by the above-mentioned rule of Jordan form
transformation and Theorem 2 of \cite{andsok1}). Statements 3(d) and
3(a) follow from the nonminimizability of $r_\mp$, from Lemma 8 and
from the fact that, by construction, $r_\mp$ intertwine the
Hamiltonians whose ground states are situated above $\lambda_1$.

Coefficients of $q_N^\mp$ are real-valued since coefficients of all
operators contained in the obtained factorizations of $q_N^\mp
P_\pm(h^\pm)$ are real-valued, as well as coefficients of
$P_\pm(h^\pm)$. Theorem 2 is proved.\\

\noindent {\bf Theorem 3} (on complete reducibility of
nonminimizable intertwining operators).

\noindent{\it Assume that the
following conditions are satisfied:

\renewcommand{\labelenumi}{\rm{(\theenumi)}}
\begin{enumerate}

\item $h^+=-\partial^2+V_1(x)$, $V_1(x)\in K$;

\item $h^-=-\partial^2+V_2(x)$, where the potential $V_2(x)$ is real-valued
and belongs to $C_{\Bbb R}$;

\item $h^+$ and $h^-$ are intertwined by a nonminimizable
differential operator $q_N^-$ of $N$th order with real-valued
coefficients from $C^2_{\Bbb R}$ , so that \be q_N^-h^+=h^-q_N^-;\ee

\item the algebraic multiplicity of $\lambda_i$, the $i$\hspace{0.3mm}th
eigenvalue of the matrix $\bf S$ for the operator $q_N^-$, is equal
to $k_i$, $i=1$, \dots, $n$, so that $k_1+\dots+k_n =N$; the set of
values $\lambda_i$ contains $M$ real values and $L$ pairs of
mutually complex conjugate ones, so that $M+2L=n$; the indices
$i=1$, \dots, $M$ correspond to real $\lambda_i$, and
$\lambda_i>\lambda_{i+1}$, $i=1$, \dots, $M-1$;

\item if $\lambda_1$ is real, then $\lambda_1\leqslant0$;

\item $E_{i\pm}$, $i=0$, $1$, $2$, \dots, is the energy of the
$i$\hspace{0.3mm}th (from below) bound state of $h^\pm$;
$K_\pm=\max\{i:\lambda_i>E_{0\pm}\}$ if $\lambda_1>E_{0\pm}$, and
$K_\pm=0$ if either $\lambda_1\leqslant E_{0\pm}$ or
${\rm{Im}}\,\lambda_1\ne0$.

\end{enumerate}

\noindent Then:

\renewcommand{\labelenumi}{\rm{(\theenumi)}}
\begin{enumerate}

\item $V_2(x)\in K$; coefficients of $q_N^-$ belong to
$C^\infty_{\Bbb R}$; $q_N^+=(q_N^-)^t$ has real-valued coefficients
from $C^\infty_{\Bbb R}$ and intertwines $h^+$ and $h^-$, so that
\be h^+q_N^+=q_N^+h^-;\ee

\item $q_N^{\mp}$ can be represented as a product of really
irreducible intertwining operators of first and second orders with
real-valued coefficients from $C^\infty_{\Bbb R}$, so that:

\renewcommand{\labelenumii}{\rm{(\theenumii)}}
\begin{enumerate}

\item potentials of all the intermediate Hamiltonians belong to
$K$;

\item the first \be J_1=\sum\limits_{i=M+1}^{M+L}k_i\ee operators
from the right in the factorization of $q_N^\mp$ under consideration
have order two and are really irreducible operators of the I type;
in addition, one can realize that the related to these operators
pairs of mutually complex conjugate eigenvalues of the matrix $\bf
S$ for the operator $q_N^-$ are ordered arbitrarily;

\item the second (from the right) group of operators in the
factorization under consideration consists of \be
J_{2\mp}=N-2J_1-2J_{3\mp},\ee operators of first order, where \be
J_{3\mp}=\Big[{1\over2}\sum\limits_{i=1}^{K_\mp}k_i\Big],\ee and

\renewcommand{\labelenumiii}{\rm{(\theenumiii)}}
\begin{enumerate}

\item if $\sum\limits_{i=1}^{K_\mp}k_i$ is even, then the eigenvalue
of the matrix $\bf S$ for the operator $q_N^-$ which corresponds to
the $l$\hspace{0.3mm}th (from the right) of these operators does not
exceed the eigenvalue related to the $(l+1)$th operator, $l=1$,
\dots, $J_{2\mp}-1$;

\item if $\sum\limits_{i=1}^{K_\mp}k_i$ is odd, then the eigenvalue
of the matrix $\bf S$ for the operator $q_N^-$ which corresponds to
the $l$\hspace{0.3mm}th (from the right) of these operators does not
exceed the eigenvalue related to the $(l+1)$th operator, $l=1$,
\dots, $J_{2\mp}-2$; $\lambda_{K_\mp}$ is the eigenvalue that
corresponds to the $(J_{2\mp}-1)$th operator and $\lambda_{K_\mp+1}$
is the eigenvalue that corresponds to the $J_{2\mp}$th operator; in
this case, the latter eigenvalue is equal to $E_{0\mp}$;

\end{enumerate}

\item the third (from the right) and the last group of operators in
the factorization under consideration consists of $J_{3\mp}$ really
irreducible operators of the II and III types, and the largest of
eigenvalues of the matrix $\bf S$ for the operator $q_N^-$ which
corresponds to the $l$\hspace{0.3mm}th of these operators (from the
right) does not exceed the smallest eigenvalue of the matrix $\bf S$
for the operator $q_N^-$ which corresponds to the $(l+1)$th of these
operators, $l=1$, \dots, $J_{3\mp}-1$.

\end{enumerate}\end{enumerate}}

\vskip0.5pc

\noindent{\bf Remark 1.} If $E_{0\mp}$ is not an eigenvalue of the
matrix $\bf S$ for the operator $q_N^-$, then
$\sum\limits_{i=1}^{K_\mp}k_i$ is even since otherwise the
eigenvalue of $q_N^\mp
q_N^\pm\equiv\prod_{i=1}^n(h^\mp-\lambda_i)^{k_i}$ at the ground
state wave function of $h^\mp$ is negative.\\

\noindent{\bf Proof.} Let us restrict ourselves by a proof for the
case of $q_N^-$ only (a proof for the case of $q_N^+$ is analogous).
The first statement follows from Lemma~1. Statement 2(b) follows
from Lemma 7. Statement 2(c) in the part that corresponds to
intertwining operators for which the eigenvalues of the matrices
$\bf S$ are situated below $E_{0-}$ follows from Lemma 8. It also
follows from Lemmas 7 and 8 that corresponding part of intermediate
Hamiltonians belongs to $K$. Thus, the proof is reduced to the case
where $L=0$ and $\lambda_M=\lambda_n\geqslant E_{0-}$, which is
assumed below.

Let us first describe the main idea of the proof. The idea is as
follows. We apply Theorem 2 to factorize the operator $q_N^-
P_+(h^+)$ into three groups of intertwining operators of first
order. Then we successively permute any operator from the right-hand
group (by Lemmas 9 and 10) with the operators of the middle group
(certainly, such an operator is changed by any permutation, but its
matrix $\bf S$ is preserved) until this operator either takes its
proper position in the middle group (if the eigenvalue of its matrix
$\bf S$ belongs to the spectrum of the matrix $\bf S$ of the
operator $q_N^-$) or pass the middle group entirely. In parallel,
one must permute operators from the left-hand group (such that
eigenvalues of their matrices $\bf S$ belong to the spectrum of the
matrix $\bf S$ of the operator $q_N^-$) with operators of the middle
group as long as they get their proper positions. In this
connection, operators of the right-hand group that pass the middle
group entirely will form, under contact with operators of the
left-hand group with the same matrices $\bf S$, differences of
eigenvalues and Hamiltonians which provide the possibility to
minimize $q_N^-P_+(h^+)$ to $q_N^-$, and thus, to get the required
factorization of~$q_N^-$ as a result.

Now we present details. We consider successively (from top to
bottom) all the energy levels of the super-Hamiltonian $H$ that are
not situated above $\lambda_1$. We start from the case of the upper
of these levels, $E_{\max}$. If $E_{\max}$ coincides with one of
eigenvalues of the matrix~$\bf S$ of the operator $q_N^-$ (so that
$E_{\max}=\lambda_i$), then we proceed as follows.

\renewcommand{\theenumi}{\alph{enumi}}
\begin{enumerate}

\item If $E_{\max}$ belongs to the spectrum of $h^+$, then we permute the
corresponding to $E_{\max}$ operator from the right-hand group of
the factorization given by Theorem 2 (obvi\-ously, this operator is
the most left in the right-hand group) with operators of the middle
group from right to left with the help of Lemma 9 until the
permutation with the most left of the operators that correspond to
$\lambda_{i+1}$.

\item If $E_{\max}$ does not belong to the spectrum of $h^+$, then cofactors
of the right-hand group contain no cofactor corresponding to
$E_{\max}$, and we do not make any permutations from right to left
with cofactors of the middle group.

\item If $E_{\max}$ belongs to the spectrum of $h^-$, then we permute the
corresponding to $E_{\max}$ operator from the left-hand group
(obviously, this operator is the most right in the left-hand group)
with operators of the middle group from left to right with the help
of Lemma 10 either until the permutation with the most right of the
operators corresponding to $\lambda_{i-1}$ (if $k_1+...+k_{i-1}$ is
even) or until the permutation after which the right-hand neighbour
of the moved operator is the most right of the operators
corresponding to $\lambda_{i-1}$ (if $k_1+...+k_{i-1}$ is odd). Let
us note that in the case under consideration, the following happens.
If two first order operators are to the right of the moved operator
from the left-hand group before a permutation and are to the left
from the moved operator after the permutation, then these two
operators form the really irreducible second order operator of the
II or III type. This is explained by the fact that both eigenvalues
of the matrix $\bf S$ of this operator are situated after the
permutation above the energy of the ground state (generated by the
moved operator from the left-hand group since an element of its
kernel is nonnormalizable at both infinities by Theorem 2 and Lemma
10); thus, both elements of a canonical basis in the kernel of the
considered operator, which are formal eigenfunctions of the proper
intermediate Hamiltonian, must have zeroes.

\item If $E_{\max}$ does not belong to the spectrum of $h^-$, then cofactors of the
left-hand group contain no cofactor corresponding to $E_{\max}$, and
we do not make any permutations from left to right with cofactors of
the middle group.

\end{enumerate}

Now we consider the case where $E_{\max}$ does not belong to the
spectrum of the matrix~$\bf S$ of the operator $q_N^-$. In this
case, there is the index $i$ such that $\lambda_i>
E_{\max}>\lambda_{i+1}$ (or $i=n=M$ and $\lambda_i>E_{\max}$). In
addition, in this case, both Hamiltonians $h^\pm$ have the level
$E_{\max}$ (see the second statement of Theorem 2); to avoid a
negative eigenvalue of the supercharges anticommutator for a wave
function of $H$ for the level $E_{\max}$, the following condition
must hold: \be k_1+...+k_i\qquad \hbox{is even}.\la{zvezda}\ee In
the considered case, the most left of the right-hand group cofactors
corresponds to $E_{\max}$. We permute this cofactor with the help of
Lemma 9 with cofactors of the middle group until the permutation
with the most left of the cofactors corresponding to
$\lambda_{i+1}$. Further permutations are accomplished with the help
of Lemma 10. In this connection, the passage of the considered
operator from the right-hand group through the entire middle group
is possible by virtue of condition \gl{zvezda}. Let us note that
after each permutation with the help of Lemma 10, the right-hand
neighbour of the moved operator is the united really irreducible
operator of the II or III type and not two separate operators of
first order. This is explained by the fact that after the
permutation, both eigenvalues of the matrix $\bf S$ of this
neighbour are situated above the energy of the ground state
(generated by the moved operator from the right-hand group since an
element of its kernel is normalizable at both infinities by Theorem
2 and Lemmas 9 and 10); thus, both elements of a canonical basis in
the kernel of the considered operator, which are formal
eigenfunctions of the proper intermediate Hamiltonian, must have
zeroes.

After passing through the middle group, the operator of the
right-hand group is located near the intertwining operator of the
left-hand group with the same matrix $\bf S$. By Theorem 2 of
\cite{andsok1} and the rule of transformation of the Jordan form of
the matrix $\bf S$ of an intertwining operator under separation from
it an intertwining operator of first order (see the proof of Lemma 1
in \cite{ansok}), the product of these operators is equal to the
difference of $E_{\max}$ and the intermediate Hamiltonian. With the
help of intertwining relations, this difference can be moved to the
bound of the considered factorization and separated.

We proceed further in the same way by induction. As a result, we
obtain the required factorization of $q_N^-$. Theorem 3 is proved.

\section*{Acknowledgments}

The author thanks A.A. Andrianov for numerous discussions and useful
remarks. This research was supported by the RFBR (project
06-01-00186-a).

\end{document}